\documentstyle[12pt]{article}
\font\twlgot =eufm10 scaled \magstep1
\font\egtgot =eufm8
\font\sevgot =eufm7
\font\twlmsb =msbm10 scaled \magstep1
\font\egtmsb =msbm8
\font\sevmsb =msbm7

\newfam\gotfam
\def\pgot{\fam\gotfam\twlgot}
\textfont\gotfam\twlgot
\scriptfont\gotfam\egtgot
\scriptscriptfont\gotfam\sevgot
\def\got{\protect\pgot}
\newfam\msbfam
\textfont\msbfam\twlmsb
\scriptfont\msbfam\egtmsb
\scriptscriptfont\msbfam\sevmsb
\def\Bbb{\protect\pBbb}
\def\pBbb{\relax\ifmmode\expandafter\Bb\else\typeout{You cann't use
Bbb in text mode}\fi}
\def\Bb #1{{\fam\msbfam\relax#1}}

\newcommand{\gA}{{\got A}}

\def\thebibliography#1{\section*{References}\list
    {[\arabic{enumi}]}{\settowidth\labelwidth{#1}\leftmargin\labelwidth
      \advance\leftmargin\labelsep
      \usecounter{enumi}}
      \def\newblock{\hskip .11em plus .33em minus .07em}
      \sloppy\clubpenalty4000\widowpenalty4000
      \sfcode`\.=1000\relax}

\newcommand{\id}{{\rm Id\,}}

\newcommand{\beq}{\begin{equation}}
\newcommand{\eeq}{\end{equation}}
\newcommand{\ben}{\begin{eqnarray}}
\newcommand{\een}{\end{eqnarray}}
\newcommand{\be}{\begin{eqnarray*}}
\newcommand{\ee}{\end{eqnarray*}}
\newcommand{\bea}{\begin{eqalph}}
\newcommand{\eea}{\end{eqalph}}

\newcommand{\cP}{{\cal P}}

\newcommand{\cL}{{\cal L}}

\newcommand{\cS}{{\cal S}}

\newcommand{\la}{\lambda}

\newcommand{\f}{\phi}

\newcommand{\m}{\mu}

\newcommand{\di}{{\rm dim\,}}

\newcommand{\w}{\wedge}

\newcounter{eqalph}
\newcounter{equationa}
\newcounter{theorem}
\newcounter{remark}
\newcounter{proposition}
\newcounter{lemma}
\newcounter{corollary}
\newcounter{definition}

\newenvironment{eqalph}{\stepcounter{equation}
\setcounter{equationa}{\value{equation}}
\setcounter{equation}{0}

\begin{eqnarray}}{\end{eqnarray}\setcounter{equation}{\value{equationa}}}

\def\theremark{\arabic{remark}}

\def\thedefinition{\arabic{definition}}

\newcommand{\mar}[1]{}

\begin{document}
\hbox{}

{\parindent=0pt

{\large\bf 50 Years of Gauge Theory. Mathematical
Aspects}\footnote{Preface to the special issue of {\it International
Journal of Geometric Methods in Modern Physics}, Web:
http://www.worldscinet.com/ijgmmp/ijgmmp.shtml}
\bigskip

{\sc G.Sardanashvily}

\medskip
Department of Theoretical Physics, Moscow State University, 117234
Mosocw, Russia

E-mail: sard@grav.phys.msu.su

Web: http://webcenter.ru/$\sim$sardan/
\medskip

{\small
{\bf Abstract}. Mathematical aspect of contemporary classical and
quantum gauge theory are sketched. 

}}
\bigskip

In 1954, C.Yang and R.Mills generalized the $U(1)$-gauge theory of
electromagnetism of W.Pauli to the non-Abelian isospin group
$SU(2)$. In 1956, R.Utiyama extended their gauge scheme to
an arbitrary finite-dimensional Lie group, including the
Lorentz group. The Yang--Mills gauge
theory provided the universal description
of interactions in Lagrangian systems with symmetries by means of gauge
potentials. In 1964-66, P.Higgs suggested
the mechanism of generating a mass of gauge bosons by means
of spontaneous symmetry breaking. Being quantized, the Yang--Mills--Higgs gauge
model is successfully applied in high energy physics in order to describe
fundamental (electromagnetic,
weak and strong) interactions and their unification
\cite{50y}. The Yang--Mills gauge theory also contributes to other
branches of physics, e.g., 
solid physics \cite{edel,mal}.
Let us briefly sketch some mathematical aspects of classical and
quantum gauge theory.

{\it Classical gauge theory}. 
In a general setting, any classical field is represented by a section of
a fiber bundle $Y\to X$ over a smooth manifold $X$.
Its dynamics is phrased in terms of jet manifolds
$J^*Y$ of $Y\to X$ \cite{book}. By gauge
transformations are meant bundle automorphisms of $Y\to X$.  

For instance, let $P\to X$ be a
principal bundle with a structure Lie group $G$. This is the case of
the Yang--Mills gauge 
model where gauge potentials are connections on
$P\to X$. Being $G$-equivariant, they are represented by
sections of the affine bundle $C=J^1P/G$. Gauge
transformations are defined as vertical (over $\id X$) automorphisms of $P\to X$.
Their infinitesimal generators are $G$-invariant 
vertical vector fields on $P$. They  
make up a Lie $C^\infty(X)$-algebra which is a projective
$C^\infty(X)$-module of finite rank. Matter fields in the
Yang--Mills gauge theory are represented by sections of a fiber
bundle $Y\to X$ associated to $P$. Spontaneous symmetry breaking takes
place if the structure group $G$ of $P$ reduces to a Lie subgroup
$H$, i.e., $P$ contains a principal subbundle whose
structure group is $H$. There is one-to-one
correspondence between the $H$-principal subbundles of $P$ and the
global sections of the quotient bundle $P/H\to X$ which  
play the role of classical Higgs fields.

Let $Y\to X$ belong to
the category of natural bundles, i.e., 
any diffeomorphism of the base $X$ gives rise to an automorphism of $Y$
treated as a gauge (general covariant) transformation. The structure group
of $Y$ is $GL(n,\Bbb R)$, $n=\di X$. The
associated principal bundle is the bundle of linear frames in the
tangent bundle $TX$ of $X$. Its structure group $GL(n,\Bbb R)$ always
reduces to the maximal compact subgroup $O(n)$. The corresponding Higgs
field is a Riemannian metric on $X$. If $X$ satisfies some topological
conditions, the structure group $GL(n,\Bbb R)$ also reduces to the
Lorentz subgroup. The corresponding Higgs field is a pseudo-Riemannian
metric on $X$. With a pseudo-Riemannian metric and connections on the
linear frame bundle, we come to metric-affine gravitation theory.

Since gauge potentials are represented by connections
on principal bundles, field theory involves the algebraic topological
characteristics \cite{eguchi}.
Namely, given a $G$-principal bundle $P\to X$, 
one  associates to any invariant
polynomial
$I_k$ on the Lie algebra of
$G$ the characteristic exterior form $\cP_{2k}(F_A)$ on $X$ 
expressed in the strength form $F_A$ of a principal
connection $A$ on $P\to X$.
It is a closed form whose
de Rham cohomology class is independent of the
choice of a connection $A$ on $P$ and, thus, is a topological
characteristic of $P$. Furthermore,
given another principal connection
$A'$ on $P$, the  global transgression formula
\be
\cP_{2k}(F_A)-\cP_{2k}(F_{A'})=dS_{2k-1}(A,A')
\ee
defines the secondary characteristic form $S_{2k-1}(A,A')$, e.g., the
Chern--Simons form. 
Characteristic forms are the important ingredient in many 
classical and quantum field models, e.g., topological field theory
\cite{birm} and anomalies \cite{bert}. Several important
characteristics, e.g., Donaldson and Seiberg--Witten invariants \cite{don,moor}
come from geometry and topology of the moduli spaces of gauge fields. 

The fiber bundle formulation of classical gauge theory has stimulated
its numerous geometric generalizations. Let us mention principal
superconnections \cite{bart}, quantum principle bundles
\cite{majid2,calow} and non-commutative gauge theory \cite{poly}.

{\it Odd fields}.  In order to be quantized, classical gauge theory
should be extended to odd fields and their jets. For this purpose, one usually calls
into play fiber bundles over 
supermanifolds 
\cite{mont,franc}. 
We describe odd variables on a smooth manifold
$X$ as generating elements of the structure ring of a
graded manifold whose body is
$X$ \cite{mpla}. This definition 
reproduces  the heuristic notion of jets of ghosts
in the BRST theory \cite{barn}. 
Recall that, by virtue of
Batchelor's theorem, any graded manifold
with a body $X$ is isomorphic to the one whose
structure sheaf 
$\gA_Q$ is formed by germs of sections of the exterior product 
$\w Q^*$, where $Q^*$ is the dual of some vector
bundle $Q\to X$. In field models, a vector bundle $Q$ is
usually given from the beginning. Therefore, it suffices to consider graded
manifolds
$(X,\gA_Q)$ where 
Batchelor's isomorphism holds. 
Accordingly, $r$-order jets of odd fields are defined as
generating elements of the structure ring of the graded
manifold
$(X,\gA_{J^rQ})$ constructed from the jet bundle $J^rQ\to X$ of
$Q$. Thus, we have 
a differential bigraded algebra $\cS^*$ over $C^\infty(X)$ whose local 
basis consists of even and odd variables $s^i$, their jets
$s^i_{\la_1\ldots \la_k}$ and the graded exterior
forms $dx^\la$, $ds^i$, $ds^i_{\la_1\ldots \la_k}$. 
The algebra $\cS^*$ is split into the
variational bicomplex. For instance, Lagrangians are represented by
horizontal densities $L=\cL d^nx$.
 
{\it Gauge transformations}. As was mentioned above, infinitesimal gauge
transformations in the Yang--Mills gauge theory make up a Lie
$C^\infty(X)$-algebra which is a finite projective
$C^\infty(X)$-module. In gauge models on natural bundles over $X$, the gauge
algebra is the infinite-dimensional Lie $\Bbb R$-algebra
of vector fields on $X$. There are gauge models where parameters of
gauge transformations depend on field variables, and they
make up a certain sh-Lie algebra \cite{fulp02}. In a general setting, any
derivation of the above-mentioned algebra $\cS^*$ which preserves its
contact ideal  can be treated as an
infinitesimal gauge transformation. This is also the case of
BRST transformations. Since the BRST operator is nilpotent, different
types of BRST cohomology are studied \cite{barn}.

{\it Quantization}. The Faddeev--Popov quantization of the
Yang--Mills gauge theory in the framework of the path integration
formulation of perturbative quantum field theory
is generally accepted \cite{fad}. The action functional in the integrand 
of the generating functional contains the gauge fixing term and the
term of ghosts which makes this action functional BRST-invariant.
The Batalin--Vilkovisky antifield quantization is applied to a wider
class of gauge systems, e.g., whose gauge algebra is reducible or it
closes only on-shell \cite{bat,gom}. The BV quantization enables one to
construct a
BRST-invariant action functional, but does not automatically provide
the path integration measure. In non-perturbative quantum field theory,
the generating functional of Euclidean Green functions of a field
algebra modelled over a nuclear space $F$ can be represented as the Fourier
transform of some positive measure $\m$ on the topological dual $F'$ of $F$ 
\cite{sard91}. For instance, in the case of a scalar field, $F$ is the
Schwartz space of  
smooth functions rapidly decreasing at infinity and $F'$ is the space of 
tempered distributions. The problem is that no measure on $F'$ can be
explicitly written, unless $F$ is finite-dimensional. Moreover, one
usually considers the Sobolev completion of the space of gauge
potentials (irreducible principle connections) which makes it into a
Hilbert, but not nuclear space.  

{\it Higgs vacuum}. The existence of the Higgs vacuum is confirmed by
experiments on the Weinberg--Salam electroweak interaction. 
In contrast with the bare Fock vacuum, the Higgs one possesses
non-zero physical characteristics and, thus, can interact with particles and
fields.  However, its true physical nature still remains unclear.
Somebody thinks of the Higgs vacuum as being a {\it sui generis}
condensate. At present, there is no mathematical model of the Higgs
vacuum, unless the 
fact that measures $\m(\f)$ and $\m(\f +{\rm const.})$
on the dual $F'$ of an infinite-dimensional nuclear space
are never equivalent. Therefore, the Higgs vacuum seems to be the most
intricate target of contemporary theoretical physics.

\end{document}